\documentclass[conference,10pt]{IEEEtran}

\usepackage[english]{babel}
\usepackage[T1]{fontenc}
\usepackage[utf8]{inputenc}
\usepackage{amsmath, amssymb, amsfonts, bm}
\usepackage{graphicx}
\usepackage{cite}
\usepackage{url}
\usepackage{booktabs}
\usepackage{algorithm, algpseudocode}
\usepackage{siunitx}
\usepackage[acronym,nonumberlist]{glossaries}
\usepackage{xcolor}
\usepackage{tikz}
\usetikzlibrary{matrix,calc, arrows.meta, positioning, angles, quotes, decorations.pathmorphing, shapes.geometric}
\usepackage{empheq} 
\usepackage[hidelinks, bookmarks=false]{hyperref} 
\usepackage{subcaption}

\usepackage[hidelinks, bookmarks=false]{hyperref}

\usepackage{soul}  
\glsdisablehyper
\newacronym{ISAC}{ISAC}{Integrated Sensing and Communication}
\newacronym{LEO}{LEO}{Low Earth Orbit}
\newacronym{OTFS}{OTFS}{Orthogonal Time Frequency Space}
\newacronym{RSMA}{RSMA}{Rate-Splitting Multiple Access}
\newacronym{DD}{DD}{Delay-Doppler}
\newacronym{SINR}{SINR}{Signal-to-Interference-plus-Noise Ratio}
\newacronym{CRB}{CRB}{Cramér-Rao Bound}
\newacronym{LMMSE}{LMMSE}{Linear Minimum Mean Square Error}
\newacronym{CSI}{CSI}{Channel State Information}
\newacronym{ICSI}{ICSI}{Imperfect Channel State Information} 
\newacronym{ISIC}{ISIC}{Imperfect Successive Interference Cancellation} 
\newacronym{QoS}{QoS}{Quality of Service}
\newacronym{AWGN}{AWGN}{Additive White Gaussian Noise}
\newacronym{LOS}{LOS}{Line-of-Sight}
\newacronym{MP}{MP}{Message Passing}
\newacronym{BP}{BP}{Belief Propagation}
\newacronym{FIM}{FIM}{Fisher Information Matrix}
\newacronym{GA}{GA}{Genetic Algorithm}
\newacronym{MIMO}{MIMO}{Multiple-Input Multiple-Output}
\newacronym{MU}{MU}{Multi-User}
\newacronym{MISO}{MISO}{Multiple-Input Single-Output} 
\newacronym{OFDM}{OFDM}{Orthogonal Frequency Division Multiplexing}
\newacronym{SAGIN}{SAGIN}{Space-Air-Ground Integrated Network}
\newacronym{NTN}{NTN}{Non-Terrestrial Network}
\newacronym{SISO}{SISO}{Single-Input Single-Output}
\newacronym{SDMA}{SDMA}{Space-Division Multiple Access}
\newacronym{NOMA}{NOMA}{Non-Orthogonal Multiple Access}
\newacronym{SP}{SP}{Signal Processing} 
\newacronym{TF}{TF}{Time-Frequency} 

\newcommand{\herm}{H}
\newcommand{\pr}{\partial}

\def\BibTeX{{\rm B\kern-.05em{\sc i\kern-.025em b}\kern-.08em
    T\kern-.1667em\lower.7ex\hbox{E}\kern-.125emX}}

\begin{document}

\title{Refined Metrics, Sensing Limits, and Resource Allocation in OTFS-RSMA LEO ISAC}

\author{\IEEEauthorblockN{{Bruno Felipe Costa} and {Taufik Abrão}}
\IEEEauthorblockA{\textit{Department of Electrical Engineering},\,\,
\textit{State University of Londrina} (UEL),\quad
Londrina, PR, Brazil \\
bruno.uel.felipe@gmail.com\qquad taufik@uel.br}
}

\maketitle
\glsresetall 

\begin{abstract}
This paper develops an integrated \gls{OTFS}-\gls{RSMA} framework employing advanced \gls{SP} techniques tailored for this demanding environment. We derive refined communication performance metrics, specifically \gls{SINR} expressions capturing the practical effects of \gls{ICSI} and \gls{ISIC}. Moreover, fundamental sensing limits are established via \gls{CRB} derivation incorporating parameter-dependent echo gain, linking waveform \gls{SP} properties to estimation accuracy. The resource allocation is formulated as a non-convex 
optimization problem aiming for Max-Min Fairness under constraints derived from these \gls{SP} metrics. Illustrative results, obtained via \gls{GA} optimization, crucially demonstrate that the proposed \gls{RSMA} scheme uniquely enables the simultaneous satisfaction of stringent communication and sensing constraints metrics, a capability not achieved by conventional \gls{SDMA}. Such results {highlight the efficacy of integrated OTFS-RSMA precoding and optimization approach} for designing robust and feasible \gls{LEO}-\gls{ISAC} systems.
\end{abstract}


\glsresetall 

\section{Introduction}
\label{sec:introduction}

Next-generation wireless systems, including 6G, envision \gls{SAGIN} architectures providing ubiquitous connectivity and environmental awareness. \Gls{ISAC} is a key enabler, enhancing spectral efficiency by using dual-function waveforms \cite{Shtaiwi2024}. \Gls{LEO} satellite systems are attractive platforms for global \gls{ISAC} deployment \cite{Park2024}, but introduce significant \gls{SP} challenges. The high satellite velocity causes large Doppler shifts, degrading conventional modulations like \gls{OFDM} \cite{Devarajalu2023}. Furthermore, managing multiuser interference and ensuring reliable sensing under power constraints and dynamic channel conditions requires sophisticated \gls{SP} solutions.

\Gls{OTFS} modulation emerges as a powerful \gls{SP} technique to combat high Doppler \cite{Hadani2017}. By operating in the \gls{DD} domain, \gls{OTFS} transforms the doubly-selective channel into a quasi-static representation, offering inherent resilience \cite{Liu2025}. This DD domain operation also aligns naturally with radar parameters (range/delay, velocity/Doppler), making \gls{OTFS} suitable for \gls{ISAC} \cite{Shtaiwi2024}. Key \gls{SP} challenges in \gls{OTFS} include efficient channel estimation \cite{Kumar2025}, robust data detection (e.g., using \gls{MP}/\gls{BP} algorithms exploiting channel sparsity \cite{Gaudio2020}), and managing inherent \gls{DD} interference \cite{Liu2025}.

\begin{figure}[!t] 
\centering
\includegraphics[width=0.9\columnwidth]{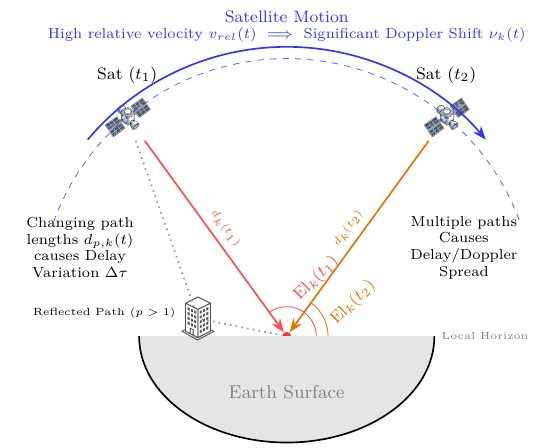} 
\caption{\small LEO motion impact on DD channel: Time-varying geometry causes 
Doppler shifts, delay variations, and multipath spread.}
\label{fig:leo_dynamics_channel_spawc} 
\vspace{-4mm} 
\end{figure}

To manage multiuser interference, \Gls{RSMA} provides a flexible \gls{SP} framework \cite{Xu2021}. By splitting messages and employing advanced precoding, \gls{RSMA} offers robust interference control compared to \gls{SDMA} or \gls{NOMA}, especially under practical impairments like \gls{ICSI} \cite{Cerna2021}. Integrating the Doppler resilience of \gls{OTFS} with the interference management flexibility of \gls{RSMA} is promising for \gls{LEO} \gls{ISAC} \cite{Huai2025, Liu2025a}. However, the core \gls{SP} task lies in designing resource allocation strategies that optimally balance communication \gls{QoS} (considering fairness \cite{Joudeh2017}), sensing precision, and power efficiency under realistic channel conditions and receiver imperfections.

This paper delves into the advanced \gls{SP} aspects of such an integrated \gls{OTFS}-\gls{RSMA} \gls{LEO} \gls{ISAC} system. The
main contributions of this work are {fivefold}:
\textbf{\textit{i}}) A concise system model incorporating advanced signal processing techniques, namely \gls{OTFS} and \gls{RSMA}, tailored for realistic \gls{LEO} channel characteristics (including \gls{DD} sparsity) and practical impairment models (\gls{ICSI}, and \gls{ISIC}). 
\textbf{\textit{ii}}) 
Derivation of refined communication performance metrics (\gls{SINR}) using \gls{LMMSE} analysis, explicitly accounting for \gls{ICSI} and \gls{ISIC} effects crucial for robust \gls{SP} system design. 
\textbf{\textit{iii}}) 
Derivation of the fundamental sensing performance limits via (\gls{CRB}) based on the \gls{FIM}, considering parameter-dependent echo gain variations and explicitly linking estimation accuracy to the second-order moments of the Delay-Doppler energy distribution of waveforms, determined by the precoders $\boldsymbol{\mathcal{P}}$.  
\textbf{\textit{iv}}) Formulation and analysis of a fairness-aware (Max-Min) resource allocation problem from an \gls{SP} perspective, highlighting its non-convex nature and discussing potential \gls{SP}-centric optimization approaches.
\textbf{\textit{v}}) Numerical demonstration, by deploying evolutionary heuristic GA optimization, in which the integrated \gls{RSMA} strategy uniquely enables the simultaneous satisfaction of stringent communication (common rate \gls{QoS}) and sensing (\gls{CRB} \gls{QoS}) constraints, showcasing a feasibility advantage over conventional \gls{SDMA} within the {proposed 
OTFS-RSMA LEO ISAC networks}. 



\section{System Model: Signal Processing Perspective}
\label{sec:system_model}

We consider a downlink \gls{MU}-\gls{MISO} \gls{ISAC} system where a \gls{LEO} satellite ($N_t$ antennas) serves $K$ single-antenna users using \gls{OTFS} modulation and \gls{RSMA}. We focus on the core \gls{SP} equations and models.

\vspace{2mm}
\noindent\textbf{OTFS Modulation Principles}.
%
\gls{OTFS} maps information symbols $x[l, m]$ onto an $M \times N$ \gls{DD} grid ($l$: delay, $m$: Doppler). These are transformed to the \gls{TF} domain via the Inverse Symplectic Finite Fourier Transform (ISFFT), yielding intermediate TF symbols denoted as $X[n, i]$ (where $n$ is the time index, $0 \le n < N$, and $i$ is the frequency index, $0 \le i < M$), and then mapped to the time domain via the Heisenberg transform \cite{Hadani2017}. The receiver reverses this using the Wigner transform and SFFT. Its key \gls{SP} advantage for \gls{LEO} {satellite networks} is transforming the time-varying channel into a quasi-static \gls{DD} representation, mitigating Doppler effects \cite{Liu2025}.

\vspace{2mm}
\noindent\textbf{RSMA Strategy}.
%
\gls{RSMA} splits user $k$'s message into common $s_c$ ($\mathbb{E}[|s_c|^2=1$) and private $s_{p,k}$ ($\mathbb{E}[|s_{p,k}|^2=1$) symbols. Linear precoding in the \gls{DD} domain provides flexible interference management, a core \gls{SP} task \cite{Mao2022}.

\vspace{2mm}
\noindent\textbf{LEO Channel and DD Representation}. 
%
The physical channel between satellite antenna $n_t$ and user $k$ is time-varying, characterized by a \gls{DD} impulse response $h_{k,n_t}(\tau, \nu; t)$ comprising $P$ paths \cite{Hadani2017}:
\begin{equation}
\label{eq:spawc_dd_impulse} 
\begin{split}
h_{k,n_t}(\tau, \nu; t) = \sum_{p=1}^{P} \alpha_{p,k,n_t}&(t) e^{j\psi_{p,k,n_t}(t)} \\ 
&\!\!\!\!\!\!\!\!\times \delta(\tau - \tau_{p,k}(t)) \delta(\nu - \nu_{p,k}(t)). 
\end{split}
\end{equation}
Parameters $(\alpha, \psi, \tau, \nu)$ depend on orbital dynamics, geometry, and fading \cite{3GPP_TR_38_811, ZiangLiu2024}. This physical channel results in an effective $N_{dd} \times N_{dd}$ \gls{DD} channel matrix $\mathbf{H}_k(t)$ relating transmitted and received \gls{DD} symbols. Crucially for \gls{SP}, due to the typically small number of paths $P \ll MN$ in satellite links, $\mathbf{H}_k(t)$ is often sparse, 
which can be exploited by advanced receivers.


\section{Advanced Performance Metrics: Derivations}
\label{sec:perf_metrics}

Accurate assessment and optimization require precise \gls{SP} metrics that account for practical impairments. We detail the derivations for refined \gls{SINR} and \gls{CRB}.

\vspace{2mm}
\noindent\textbf{Refined SINR under ICSI and ISIC}. 
%
We derive approximate \gls{SINR} expressions for \gls{LMMSE} receivers designed with $\hat{\mathbf{H}}_k$, explicitly incorporating \gls{ICSI} (variance $\sigma_e^2$) and \gls{ISIC} (factor $\Theta_k$), following established methodologies \cite{ Mishra2022}. The filter vectors are $\mathbf{w}_{c,k}, \mathbf{w}_{p,k} \in \mathbb{C}^{1 \times N_{dd}}$.

\subsubsection{Common Stream SINR Derivation}
The desired signal power is $P_{S,c} = |\mathbf{w}_{c,k} \hat{\mathbf{H}}_k \boldsymbol{\mathcal{P}}_c|^2$. Interference comes from private streams. The effective noise includes filtered AWGN and the impact of ICSI on all signal components. The total interference-plus-noise power $D_c$ is approximated as:
\begin{equation}
    D_c \approx \sum_{j=1}^K |\mathbf{w}_{c,k} \hat{\mathbf{H}}_k \boldsymbol{\mathcal{P}}_{p,j}|^2 + \|\mathbf{w}_{c,k}\|^2 (\sigma_n^2 + \sigma_e^2 P_{\text{tot}}).
    \label{eq:spawc_Dc}
\end{equation}
The term $\sigma_e^2 P_{\text{tot}}$ accounts for the variance introduced by the channel error $\mathbf{E}_k$ acting on the total transmitted signal power $P_{\text{tot}}$, projected onto the filter space.
The refined SINR for the common stream at user $k$ is:
\begin{empheq}{equation}
\small
\mathrm{SINR}_{c,k}^{\text{(ref)}} \approx \frac{| \mathbf{w}_{c,k} \hat{\mathbf{H}}_k \boldsymbol{\mathcal{P}}_c |^2 }{\sum_{j=1}^K | \mathbf{w}_{c,k} \hat{\mathbf{H}}_k \boldsymbol{\mathcal{P}}_{p,j} |^2 + \|\mathbf{w}_{c,k}\|^2 (\sigma_n^2 + \sigma_e^2 P_{\text{tot}})}
\label{eq:spawc_sinr_c_final}
\end{empheq}

\subsubsection{Private Stream SINR Derivation}

To facilitate the analysis and presentation of the private stream SINR, we first define its constituent signal, interference, and noise power components. Let $P_{S, p, k}$ denote the desired signal power for user $k$'s private stream, $I_{p, k}^{\text{(inter)}}$ the interference power from other users' private streams, $I_{p, k}^{\text{(res)}}$ the residual interference power from the common stream due to imperfect SIC (ISIC) with factor $\Theta_k$, and $N_{p, k}^{\text{(eff)}}$ the effective noise power incorporating AWGN and the impact of imperfect CSI (ICSI) with error variance $\sigma_e^2$. These components are given by:
\begin{align}
    P_{S, p, k} &= | \mathbf{w}_{p,k} \hat{\mathbf{H}}_k \boldsymbol{\mathcal{P}}_{p,k} |^2 \label{eq:spawc_sinr_p_sig_power} \\
    I_{p, k}^{\text{(inter)}} &= \sum_{j \neq k} | \mathbf{w}_{p,k} \hat{\mathbf{H}}_k \boldsymbol{\mathcal{P}}_{p,j} |^2 \label{eq:spawc_sinr_p_inter_interf} \\
    I_{p, k}^{\text{(res)}} &= \Theta_k |\mathbf{w}_{p,k} \hat{\mathbf{H}}_k \boldsymbol{\mathcal{P}}_c|^2 \label{eq:spawc_sinr_p_resid_interf} \\
    N_{p, k}^{\text{(eff)}} &= \|\mathbf{w}_{p,k}\|^2 (\sigma_n^2 + \sigma_e^2 P_{\text{tot}}) \label{eq:spawc_sinr_p_eff_noise}
\end{align}

Using these definitions, the refined SINR for the private stream of user $k$ can be expressed compactly as:

\begin{equation} \label{eq:spawc_sinr_p_compact}
    \mathrm{SINR}_{p,k}^{\text{(ref)}} \approx \frac{P_{S, p, k}}{I_{p, k}^{\text{(inter)}} + I_{p, k}^{\text{(res)}} + N_{p, k}^{\text{(eff)}}}.
\end{equation}

These refined metrics, namely $\mathrm{SINR}_{c,k}^{\text{(ref)}}$ \eqref{eq:spawc_sinr_c_final} and $\mathrm{SINR}_{p,k}^{\text{(ref)}}$ \eqref{eq:spawc_sinr_p_compact}, provide a more accurate basis for system design under practical SP impairments.

\vspace{2mm}
\noindent\textbf{Sensing CRB with Variable Gain}. 
We derive the \gls{FIM} and \gls{CRB} for estimating target delay ($\tau_T$) and Doppler ($\nu_T$) when the complex echo path gain $\alpha$ varies with delay, $\alpha = \alpha(\tau_T)$. This requires differentiating the mean received echo signal $\boldsymbol{\mu}(\tau_T, \nu_T)$ w.r.t. the parameters \cite{Kay1993}. The noise variance is $\sigma_{\text{echo}}^2 = \sigma^2$. This follows standard FIM calculation procedures adapted for parameter-dependent amplitudes and the OTFS structure \cite{Wu2024}.

\subsubsection{Mean Echo Signal and Derivatives}
The mean received signal in the \gls{DD} domain, element-wise, is \cite{costa2025otfs}:
\begin{equation}
\mu[l,k] = \frac{\alpha(\tau_T) \beta_T}{\sqrt{MN}} \sum_{n,i} X[n,i] e^{j\phi_{n,i,l,k}},
\end{equation}
where $\beta_T$ is target reflectivity, $X[n,i]$ are transmitted TF symbols, and $\phi_{n,i,l,k} = 2\pi [ n( \nu_T T - l/N) - i(\tau_T \Delta f - k/M) ]$. The derivatives are showed in  \cite{costa2025otfs}, using $\pr \alpha / \pr \tau_T = -2\alpha(\tau_T)/\tau_T$):

\begin{align}
\small
 \frac{\pr \mu[l,k]}{\pr \nu_T} &= \frac{j2\pi T \alpha(\tau_T) \beta_T}{\sqrt{MN}} \sum_{n,i} n X[n,i] e^{j\phi_{n,i,l,k}} \label{eq:spawc_dmu_dnu} \\
 \small
 \frac{\pr \mu[l,k]}{\pr \tau_T} &= \frac{-2}{\tau_T} \mu[l,k] - \frac{j2\pi \Delta f \alpha(\tau_T) \beta_T}{\sqrt{MN}} \sum_{n,i} i X[n,i] e^{j\phi_{n,i,l,k}} \label{eq:spawc_dmu_dtau}
\end{align}

Let the vectors containing these derivatives be $\mathbf{d}_{\nu}$ and $\mathbf{d}_{\tau} = \mathbf{d}_{\text{gain}} + \mathbf{d}_{\text{phase},\tau}$.

\subsubsection{FIM Element Calculation}
The FIM elements are $[\mathbf{I}^{(\text{exact})}]_{i,j} = \frac{2}{\sigma^2} \text{Re}\{ (\mathbf{d}_{i})^{\herm} \mathbf{d}_{j} \}$. Define intermediate quantities  \cite{costa2025otfs}:
\begin{align*}
S_p &\triangleq \sum_{l,k} \left| \sum_{n,i} p X[n,i] e^{j\phi_{n,i,l,k}} \right|^2, \text{ with } p=n, i\\
C_{\tau\nu} &\triangleq \text{Re}\{ \mathbf{d}_{\text{phase},\tau}^{\herm} \mathbf{d}_{\nu} \} \quad
C_{\mu\tau} \triangleq \text{Re}\{ \boldsymbol{\mu}^{\herm} \mathbf{d}_{\text{phase},\tau} \} \\
C_{\mu\nu} &\triangleq \text{Re}\{ \boldsymbol{\mu}^{\herm} \mathbf{d}_{\nu} \} \quad \hspace{6mm}
P_{\mu} \triangleq \|\boldsymbol{\mu}\|^2
\end{align*}
Note that in the sums defining $S_n$ and $S_i$ above, the indices $n \in \{0,\dots,N-1\}$ and $i \in \{0,\dots,M-1\}$ refer to the time and frequency grid points, respectively, of the transmitted TF symbols $X[n,i]$ (derived via ISFFT in Sec. II-A).

{By adopting $|\alpha|^2 = |\alpha(\tau_T)|^2, |\beta|^2 = |\beta_T|^2$ for simplicity and compactness, the FIM elements become \cite{costa2025otfs}:} 
%
%
\begin{align}
I_{\nu\nu}^{(\text{exact})} &= \frac{2 (2\pi T)^2 |\alpha|^2 |\beta|^2}{MN \sigma^2} S_n \label{eq:spawc_Inunu_final} \\
I_{\tau\tau}^{(\text{exact})} &= \frac{8 P_{\mu}}{\sigma^2 \tau_T^2} + \frac{2(2\pi \Delta f)^2 |\alpha|^2 |\beta|^2}{MN \sigma^2} S_i - \frac{8 C_{\mu\tau}}{\sigma^2 \tau_T} \label{eq:spawc_Itautau_final} \\
I_{\tau\nu}^{(\text{exact})} &= \frac{2}{\sigma^2} C_{\tau\nu} - \frac{4}{\sigma^2 \tau_T} C_{\mu\nu} \label{eq:spawc_Itaunu_final}
\end{align}

\subsubsection{CRB Inversion}
The CRBs are the diagonal elements of $(\mathbf{I}^{(\text{exact})})^{-1}$.
$\det(\mathbf{I}^{(\text{exact})}) = I_{\tau\tau}^{(\text{exact})} I_{\nu\nu}^{(\text{exact})} - (I_{\tau\nu}^{(\text{exact})})^2$.
\begin{empheq}{equation} \label{eq:spawc_CRB_tau_explicit_final}
\small
\mathrm{CRB}(\tau_T)^{(\text{exact})} \!=\! \frac{I_{\nu\nu}^{(\text{exact})}}{\det(\mathbf{I}^{(\text{exact})})} \!=\! \frac{
    \frac{2 (2\pi T)^2 |\alpha|^2 |\beta|^2}{MN \sigma^2} S_n
}{
    \det(\mathbf{I}^{(\text{exact})}) 
}
\end{empheq}

\begin{empheq}{equation} 
\label{eq:spawc_CRB_nu_explicit_final_split} 
\small 
\begin{split}
\mathrm{CRB}(\nu_T)^{(\text{exact})} &= \frac{I_{\tau\tau}^{(\text{exact})}}{\det(\mathbf{I}^{(\text{exact})})} \\ 
&= \frac{ 
\frac{8 P_{\mu}}{\sigma^2 \tau_T^2} + \frac{2(2\pi \Delta f)^2 |\alpha|^2 |\beta|^2}{MN \sigma^2} S_i - \frac{8 C_{\mu\tau}}{\sigma^2 \tau_T}
}{
\det(\mathbf{I}^{(\text{exact})}) 
}. 
\end{split}
\end{empheq}

where the determinant in the denominators is calculated using Eqs. \eqref{eq:spawc_Inunu_final}-\eqref{eq:spawc_Itaunu_final}. These exact CRBs quantify the best achievable sensing accuracy under the variable gain model, highlighting the dependence on waveform properties ($S_n, S_i, C_{\tau\nu}$, etc.) derived from the precoders $\boldsymbol{\mathcal{P}}$.







\vspace{2mm}
\noindent\textbf{Advanced MP/BP Detection}. 
%

Advanced \gls{SP} detectors based on \gls{MP}/\gls{BP} can offer significant performance gains for \gls{OTFS}, effectively exploiting the inherent \gls{DD} channel sparsity \cite{Gaudio2020}. Although \gls{LMMSE}-based metrics are utilized herein for tractable resource allocation design (Section \ref{sec:optimization_sp}), evaluating the performance gain from MP/BP detectors remains an important consideration for this framework.

\section{Resource Allocation Optimization} 
\label{sec:optimization_sp}

The design of the \gls{RSMA} precoders $\boldsymbol{\mathcal{P}} = \{\boldsymbol{\mathcal{P}}_c, \boldsymbol{\mathcal{P}}_{p,1}, \dots, \boldsymbol{\mathcal{P}}_{p,K}\}$ is fundamentally an \gls{SP} optimization problem. We aim to shape the transmitted signals to ensure fairness and meet dual communication/sensing \gls{QoS} requirements.

\vspace{2mm}
\noindent\textbf{Max-Min Fairness Optimization Problem}. We adopt the Max-Min Fairness objective, maximizing the minimum private rate $R_{min}$ among users, subject to common rate, sensing accuracy (CRB), and power constraints. The problem is:
\begin{subequations}
\label{eq:spawc_optimization_problem_maxmin}
\begin{align}
    \underset{\boldsymbol{\mathcal{P}}, R_{min}}{\text{maximize}} \quad & R_{min} \label{eq:spawc_obj_maxmin} \\
    \text{subject to} \quad
    & R_{p,k}(\boldsymbol{\mathcal{P}}, \hat{\mathbf{H}}_k^*) \ge R_{min}, && \forall k \label{eq:spawc_const_rate_p} \\
    & R_{c,k}(\boldsymbol{\mathcal{P}}, \hat{\mathbf{H}}_k^*) \ge R_c^{\text{req}}, && \forall k \label{eq:spawc_const_rate_c} \\
    & \mathrm{CRB}(\tau_T)(\boldsymbol{\mathcal{P}}) \le \epsilon_\tau \label{eq:spawc_const_crb_tau} \\
    & \mathrm{CRB}(\nu_T)(\boldsymbol{\mathcal{P}}) \le \epsilon_\nu \label{eq:spawc_const_crb_nu} \\
    & \|\boldsymbol{\mathcal{P}}_c\|^2 + \sum_{k=1}^K \|\boldsymbol{\mathcal{P}}_{p,k}\|^2 \le P_{\text{max}} \label{eq:spawc_const_power} \\
    & R_{min} \ge 0
\end{align}
\end{subequations}

Herein, $R_{p,k} = \log_2(1 + \mathrm{SINR}_{p,k}^{\text{(ref)}})$ using \eqref{eq:spawc_sinr_p_compact}, $R_{c,k} = \log_2(1 + \mathrm{SINR}_{c,k}^{\text{(ref)}})$ using \eqref{eq:spawc_sinr_p_compact}, and CRBs are from \eqref{eq:spawc_CRB_tau_explicit_final}, (16).  
These constraints link the \gls{SP} precoder design directly to system-level QoS.

\vspace{2mm}
\noindent\textbf{Problem Analysis and Solution Approaches}. 
Problem \eqref{eq:spawc_optimization_problem_maxmin} is non-convex due to the fractional SINR expressions in rate constraints \eqref{eq:spawc_const_rate_p}-\eqref{eq:spawc_const_rate_c} and the complex dependence of CRB constraints \eqref{eq:spawc_const_crb_tau}-\eqref{eq:spawc_const_crb_nu} on $\boldsymbol{\mathcal{P}}$ (via waveform moments $S_n, S_i,$ etc.). The high dimensionality ($N_{dd}(K+1)$ variables) adds further complexity.

Finding globally optimal solutions is generally intractable. \gls{SP}-related optimization techniques are required. Potential approaches include:
\begin{itemize}
    \item \textbf{Iterative Methods:} Techniques like Successive Convex Approximation (SCA) or Block Coordinate Descent (BCD) can find locally optimal solutions by iteratively solving approximated convex subproblems or optimizing subsets of variables.
    \item \textbf{WMMSE Approach:} The Weighted Minimum Mean Square Error algorithm is widely used for non-convex rate maximization problems in wireless communications \cite{Christensen2008}, often providing good performance by iteratively optimizing precoders and MMSE weights. Adapting WMMSE to handle the coupled ISAC constraints (SINR and CRB) is a relevant direction.
    \item \textbf{Fractional Programming (FP):} Techniques like Dinkelbach's algorithm can handle the fractional SINR terms.
    \item \textbf{Metaheuristics:} For initial exploration or when analytical methods are too complex, {evolutionary heuristics like} Genetic Algorithms (GA) \cite{Goldberg1989} can be {applied successfully,} 
    although scalability can be a concern.
    \item \textbf{Machine Learning:} Emerging ML-based approaches might learn resource allocation policies.
\end{itemize}
Developing efficient, specialized algorithms based on these \gls{SP} {tools and} 
principles, tailored to the {solve and implement the hybrid, integrated OTFS-RSMA ISAC design, able to deal with the DD domain and coupled SINR/CRB constraints under ICSI/ISIC, is crucial} for practical implementation and remains an important research avenue.

\vspace{-1.1mm}
\section{Illustrative Numerical Results}
\label{sec:results_sp} 
\vspace{-1.2mm}

We present numerical results to illustrate the performance and \gls{SP} trade-offs inherent to the proposed OTFS-RSMA LEO ISAC framework. Due to the non-convex nature of the Max-Min Fairness optimization problem formulated in \eqref{eq:spawc_optimization_problem_maxmin}, stemming from the fractional \gls{SINR} expressions and coupled \gls{CRB} constraints, we employ a metaheuristic approach. Specifically, a Genetic Algorithm (GA) is utilized to find the \gls{RSMA} precoders $\boldsymbol{\mathcal{P}}$ that maximize the minimum user rate ($R_{\min}$) while satisfying the communication (\gls{QoS} $R_c^{\text{req}}$) and sensing (\gls{QoS} $\epsilon_\tau, \epsilon_\nu$) constraints through a penalty-based mechanism within the GA's fitness evaluation \cite{Goldberg1989}. Performance is evaluated over $N_{mc}=3000$ Monte Carlo simulations for statistical robustness, using the refined \gls{LMMSE}-based \gls{SINR} metrics accounting for \gls{ICSI}/\gls{ISIC} (Section \ref{sec:perf_metrics}) and the \gls{CRB} derived under a variable echo gain model.

\vspace{2mm}
\noindent\textbf{Simulation Setup}
Key simulation parameters, reflecting the LEO context and \gls{SP} considerations such as the \gls{OTFS} grid dimensions ($M=4, N=8$), sparse \gls{DD} channel model, \gls{ICSI} level ({relative power:} $\sigma_e^2= -25$ dB), and \gls{ISIC} factor ($\Theta_k=0.03$), are summarized in Table \ref{tab:spawc_sim_params_updated}.

\begin{table}[!t]
\centering
\caption{\small Simulation Parameters.} 
\label{tab:spawc_sim_params_updated} 
\scriptsize 
\renewcommand{\arraystretch}{1.0} 
\begin{tabular}{@{}ll@{}}
\toprule
Parameter & Value \\
\midrule
System Config. & MU-MISO ($N_t=4$, $K=2$) \\
Modulation & OTFS ($M=4, N=8, N_{dd}=32$) \\ 
Channel Model & Sparse DD, $P=4$ paths \\ 
\quad Path Delays ($l_p$) & $[0, 1, 2, 3]$ (indices) \\ 
\quad Path Dopplers ($k_p$) & $[0, 2, -1, 3]$ (indices) \\
Optimization Alg. & Genetic Algorithm (GA) \\ 
\quad GA Max Gen. | Pop. Size & 125 | 100 \\ 
Comm. Metric & Avg. $R_{\min}$ via refined LMMSE SINR \\ 
Sensing Metric & Avg. CRB($\tau_T$), Avg. CRB($\nu_T$) via derived expr. \\ 
Imperfect CSI ($\sigma_e^2$) & Rel. Power = -25 dB \\ 
Imperfect SIC ($\Theta_k$) & $0.03$ ($\forall k$) \\ 
Total Power ($P_{\text{max}}$) & 1.0 W (0 dBW) \\
Communication SNR & 25 dB \\ 
Echo Path SNR & 10 dB \\ 
Target Sensing Params & $\tau_T=1.0\times 10^{-4}$ s, $\nu_T=4687.5$ Hz \\ 
\quad (for CRB calc.) & $\beta_T=1.0$, $\alpha(\tau_T)=0.1$ \\ 
Comm. QoS ($R_c^{\text{req}}$) & 0.1 bps/Hz \\ 
Sensing QoS ($\epsilon_\tau$) & $2.0 \times 10^{-11}$ s$^2$ \\ 
Sensing QoS ($\epsilon_\nu$) & $5.0 \times 10^{3}$ Hz$^2$ \\ 
RSMA Parameter & $\alpha \in \{0, 0.1, 0.2, 0.3, 0.5, 0.8, 1.0\}$ \\
\# Monte Carlo Frames ($N_{mc}$) & 3000 \\
\bottomrule
\end{tabular}
\vspace{-.1mm} 
\end{table}

\begin{table}[!t]
\centering
\caption{\small  Average Performance Metrics vs. $\alpha$ ($N_{mc}=3000$).}
\label{tab:consolidated_results_updated} %
\scriptsize 
\renewcommand{\arraystretch}{1.0} 
\begin{tabular}{@{}ccccc@{}} 
\toprule
\textbf{Alpha} & \textbf{Avg. $\boldsymbol{R_{\min}}$} & \textbf{Avg. CRB($\boldsymbol{\tau_T}$)} & \textbf{Avg. CRB($\boldsymbol{\nu_T}$)} & \textbf{Rc Met (\%)} \\
$\boldsymbol{(\alpha)}$ & \textbf{(bps/Hz)} & \textbf{($\boldsymbol{\times 10^{-14}}$ s$^2$)} & \textbf{($\boldsymbol{\times 10^{4}}$ Hz$^2$)} & $\boldsymbol{(\ge R_c^{req})}$\\
\midrule
0.0 & 4.884 & 563.1  & 0.497 &   0 \\ 
0.1 & 4.982 & 399.3  & 0.496 & 100 \\ 
0.2 & 4.886 & 367.6  & 0.495 & 100 \\ 
0.3 & 4.724 & 355.8  & 0.495 & 100 \\ 
0.5 & 4.278 & 369.9  & 0.495 & 100 \\ 
0.8 & 2.976 & 493.1  & 0.496 & 100 \\ 
1.0 & 0.000 & 1873.2 & 0.632 & 100 \\ 
\bottomrule
\end{tabular}
\vspace{-4mm}
\end{table}

\begin{figure}[!t]
\centering
\begin{subfigure}[b]{0.48\columnwidth}
\centering
\includegraphics[width=\linewidth]{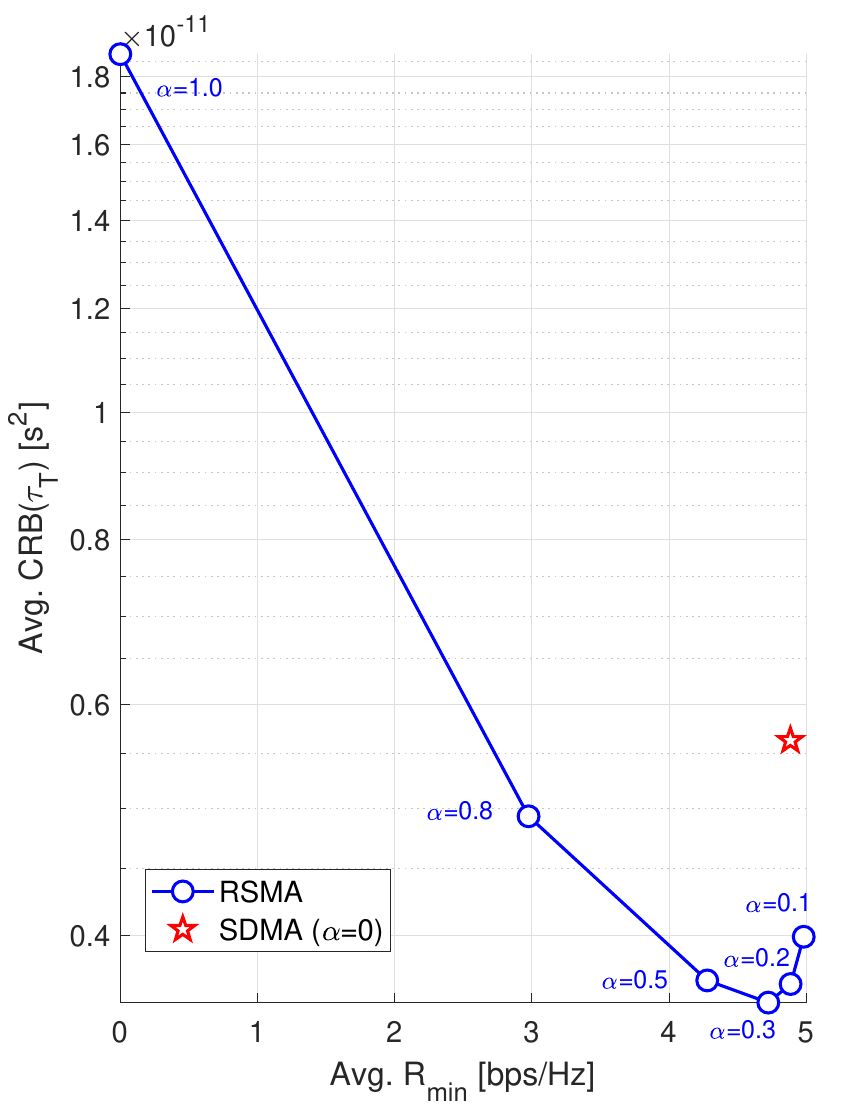}
\caption{\small  Avg. CRB($\boldsymbol{\tau_T}$) vs. Avg. $\boldsymbol{R_{\min}}$ Trade-off. 
} \label{fig:tradeoff_spawc}
\end{subfigure}
\hfill 
\begin{subfigure}[b]{0.48\columnwidth}
    \centering
    \includegraphics[width=\linewidth]{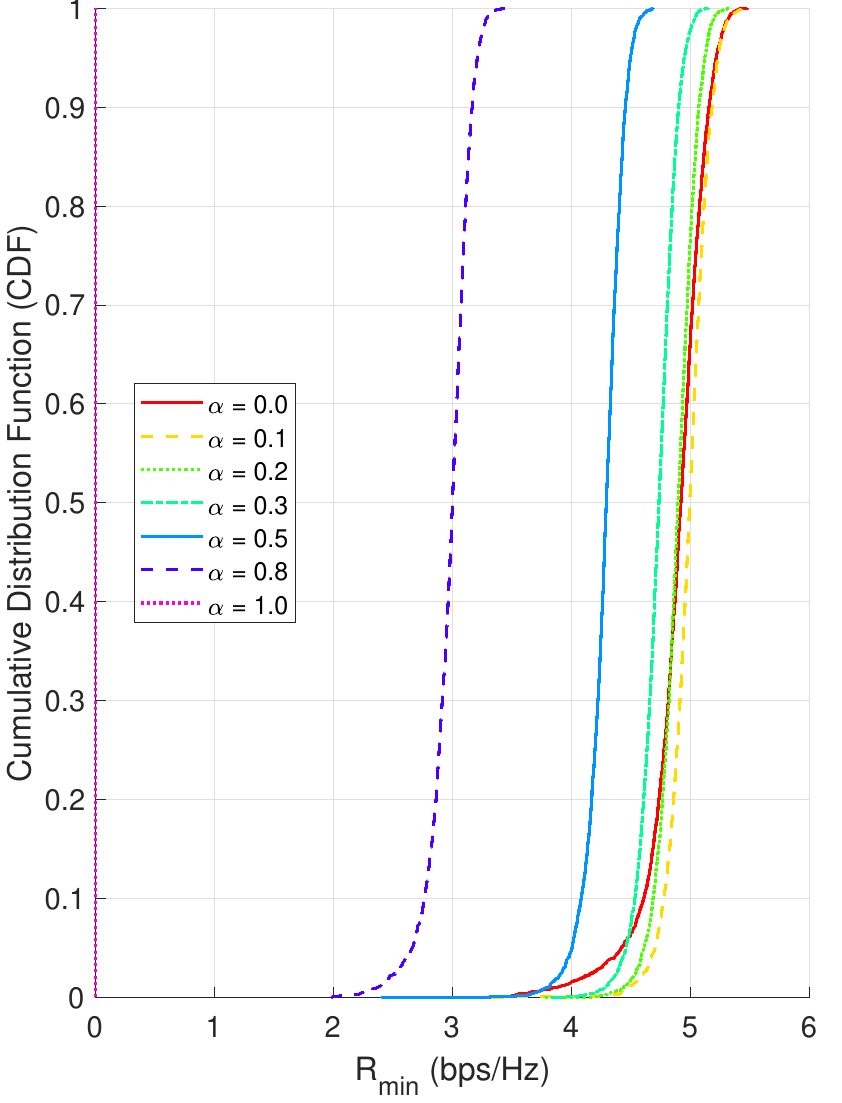}
    \caption{\small  CDF of Minimum Rate ($R_{\min}$).}
    \label{fig:cdf_rmin_spawc}
\end{subfigure}

\vspace{2mm} 

\begin{subfigure}[b]{0.48\columnwidth}
    \centering
    \includegraphics[width=\linewidth]{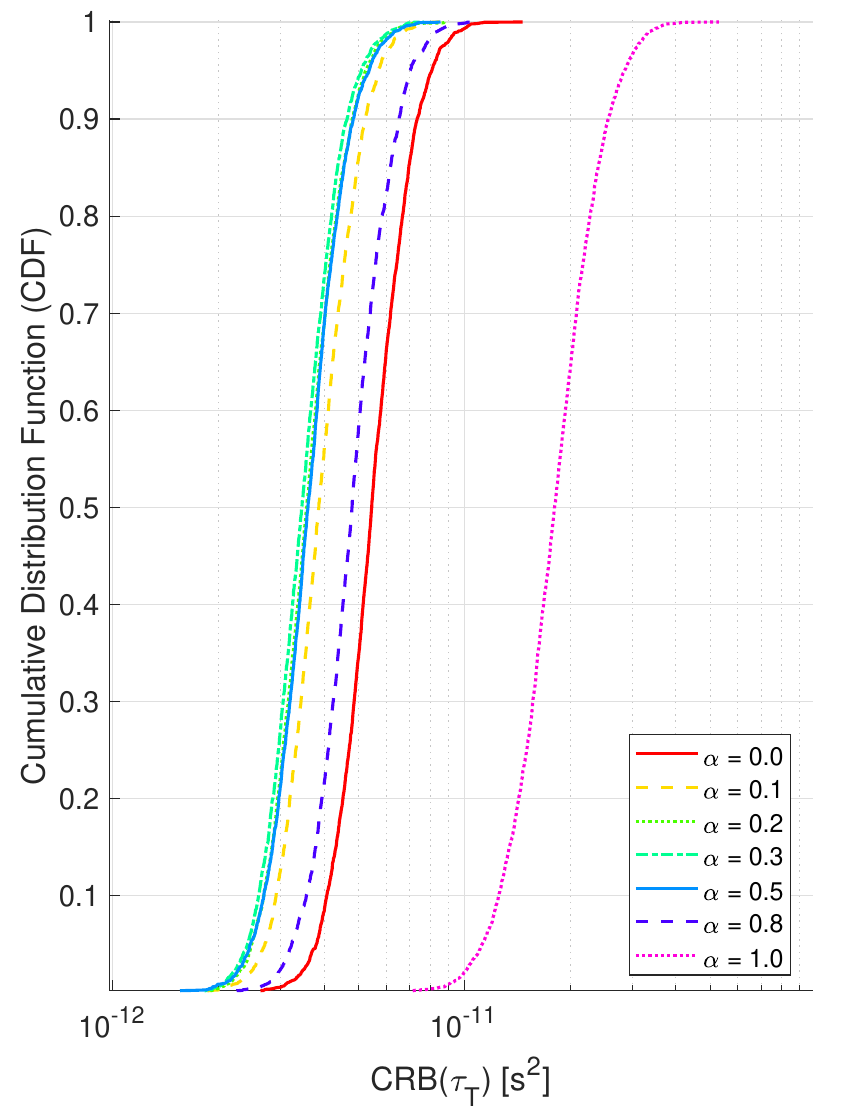}
    \caption{\small  CDF of Delay CRB (CRB($\tau_T$)).}
    \label{fig:cdf_crb_tau_spawc}
\end{subfigure}
\hfill 
\begin{subfigure}[b]{0.48\columnwidth}
    \centering
    \includegraphics[width=\linewidth]{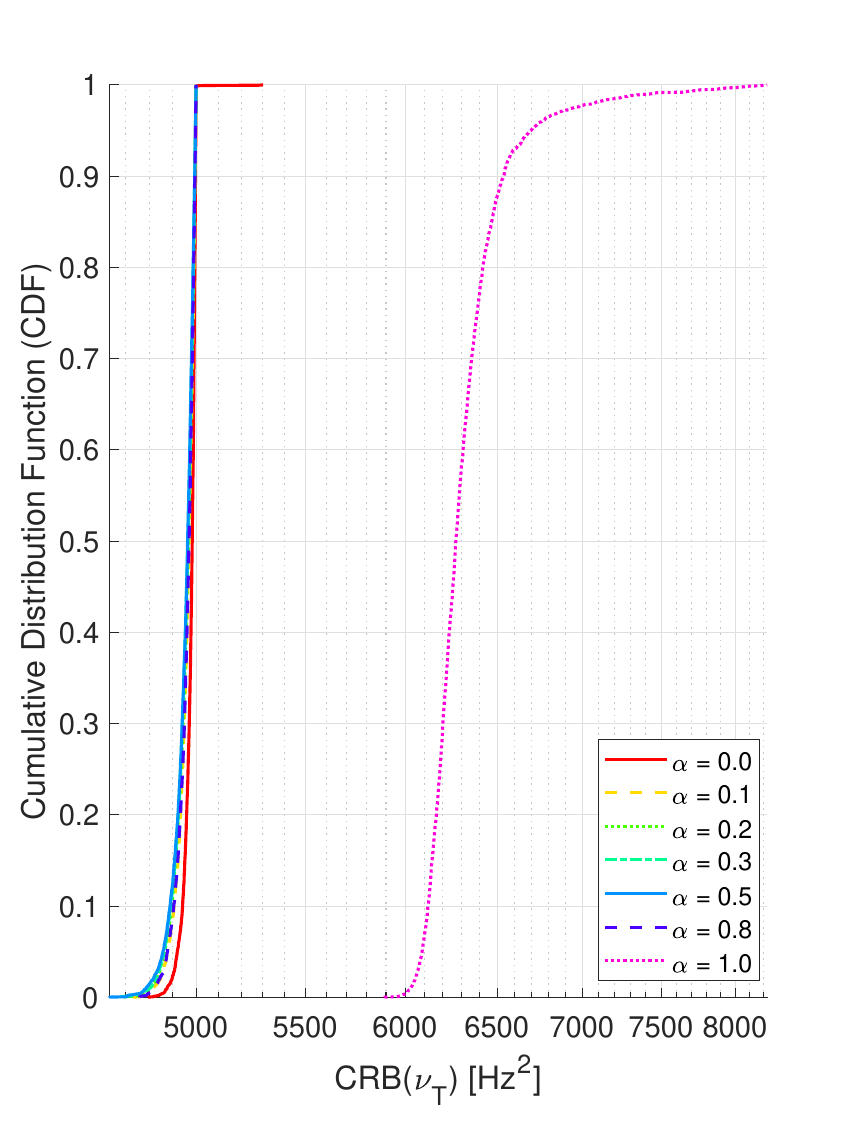} 
    \caption{\small  CDF of Doppler CRB (CRB($\nu_T$)).}
    \label{fig:cdf_crb_nu_spawc} 
\end{subfigure}

\caption{\small  Performance evaluation (Nmc=3000, M=4, N=8): (a) Fundamental ISAC trade-off. (b)-(d) CDF's illustrating performance reliability for different $\alpha$ values.}
\label{fig:spawc_results} 
\vspace{-4mm} 
\end{figure}

\vspace{2mm}
\noindent\textbf{Performance Analysis from an SP Perspective}. 
The numerical results, summarized in Fig. \ref{fig:spawc_results} and Table \ref{tab:consolidated_results_updated}, provide crucial insights into the framework's capabilities, particularly concerning \gls{SP} aspects like optimization feasibility and performance reliability under the defined constraints.

Fig. \ref{fig:spawc_results}(a) illustrates the fundamental ISAC trade-off between the average minimum user rate ($R_{\min}$) and the average delay sensing precision (CRB($\tau_T$)). As observed from Table \ref{tab:consolidated_results_updated}, the optimal average $R_{\min}$ (4.98 bps/Hz) is achieved near $\alpha=0.1$, while the best average CRB($\tau_T$) ($3.56 \times 10^{-12}$ s$^2$) occurs at $\alpha=0.3$. This confirms the ability to tune the system's focus by adjusting the \gls{RSMA} splitting factor $\alpha$.

Beyond average performance, the reliability offered by different strategies is revealed by the Cumulative Distribution Function (CDF)  plots in Figs. \ref{fig:spawc_results}(b)-(d). The steep slopes of the CDFs for $R_{\min}$, CRB($\tau_T$), and CRB($\nu_T$) in the operational RSMA region (e.g., $0.1 \le \alpha \le 0.5$) indicate low performance variance.  From an \gls{SP} perspective, this  finding demonstres that the optimized RSMA solutions are robust, achieving performance close to the average with high probability.

Crucially, the analysis highlights the limitations of \gls{SDMA} ($\alpha=0$) within the constrained optimization context. As detailed in Table \ref{tab:consolidated_results_updated}, SDMA categorically fails to meet the common rate requirement ($R_c^{\text{req}} = 0.1$ bps/Hz), achieving 0\% satisfaction ('Rc Met (\%) = 0'). In stark contrast, all evaluated \gls{RSMA} strategies ($\alpha \ge 0.1$) achieve 100\% satisfaction for this vital communication \gls{QoS} constraint, while also meeting the sensing \gls{QoS} targets (as indicated by the near-zero violation frequencies observed in the penalty analysis). This demonstrates the superior capability of \gls{RSMA}, enabled by the flexible power allocation between common and private streams, to provide the necessary degrees of freedom for the \gls{SP} optimization algorithm (GA) to find feasible solutions satisfying multiple, potentially conflicting, ISAC requirements. SDMA, lacking this flexibility, proves inadequate for this constrained multi-objective problem.

These results underscore the effectiveness of the integrated OTFS-RSMA framework, managed via \gls{SP}-based resource allocation, in providing a robust, flexible, and QoS-aware solution for \gls{LEO} \gls{ISAC} systems, particularly showcasing RSMA's advantage in scenarios with stringent and diverse service requirements.

\section{Conclusion}
\label{sec:conclusion} 

This paper focused on the advanced \gls{SP} aspects crucial for realizing \gls{ISAC} in challenging \gls{LEO} satellite environments using an integrated \gls{OTFS}-\gls{RSMA} framework. 
We discussed receiver \gls{SP} strategies, including baseline \gls{LMMSE} and advanced \gls{MP}/\gls{BP} detectors that leverage the inherent sparsity of the \gls{OTFS} \gls{DD} channel. The resource allocation task was formulated as a non-convex \gls{SP} optimization problem aiming for Max-Min Fairness, constrained by the derived SINR and CRB metrics. Numerical results, obtained via GA-based optimization solving this problem.
These results demonstrated RSMA's unique capability, in contrast to SDMA's failure, to consistently satisfy the stringent common rate communication QoS constraint ($R_c^{\text{req}}$) while simultaneously respecting the sensing performance limits (CRB constraints). This highlights the advantage of RSMA's structure in enabling the SP optimization to find feasible operating points in the constrained ISAC parameter space. Furthermore, the analysis revealed that RSMA offers a tunable communication-sensing trade-off, with CDF analysis confirming the high reliability 
achieved within the operational RSMA regimes, {\it i.e.,}, $0.1 \le \alpha \le 0.5$.



Designing efficient and scalable optimization algorithms, potentially based on iterative methods (SCA/WMMSE) or machine learning, represents a key next step to improve upon the heuristic approach and fully realize the potential of this integrated framework for \gls{LEO} \gls{ISAC}.

\vspace{-1mm}

\end{document}